\documentclass{article}

 \usepackage[preprint, nonatbib]{neurips_2026}

\usepackage[utf8]{inputenc} 
\usepackage[T1]{fontenc}    
\usepackage{hyperref}       
\usepackage{url}            
\usepackage{booktabs}       
\usepackage{amsmath}
\usepackage{amsfonts}       
\usepackage{nicefrac}       
\usepackage{microtype}      
\usepackage{xcolor}         
\usepackage[dvipsnames]{xcolor}
\usepackage{caption} 
\usepackage{subcaption} 
\usepackage{tikz}
\usetikzlibrary{quantikz2}
\usepackage{cite}
\usepackage{multirow}
\usepackage[normalem]{ulem}

\renewcommand{\ket}[1]{{| #1 \rangle}}

\newcommand{\ketbra}[2]{| #1 \rangle\!\langle #2 |}
\renewcommand{\braket}[2]{\langle #1 | #2 \rangle }

\title{Normalizing Flows for Classical Simulation of Quantum Circuits in Phase Space}
\workshoptitle{Machine Learning and the Physical Sciences}

\author{%
  Edward Jiang \\
     Quantum Information Theory Group\\
    ICFO - Institute of Photonic Sciences\\
    Barcelona Institute of Science and Technology \\ 
  \texttt{ye-duo.jiang@icfo.eu}
  \And
  Timothy Heightman \\
     Quantum Information Theory Group\\
    ICFO - Institute of Photonic Sciences\\
    \makebox[6.5cm][c]{Barcelona Institute of Science and Technology} \\
  \And
  Marcin P{\l}odzie{\'n} \\
  Instytut Fizyki Teoretycznej\\
  Wydzia\l{} Fizyki, Astronomii i Informatyki Stosowanej\\
    Uniwersytet Jagiello\'nski
}

\begin{document}

\maketitle

\begin{abstract}
    We present a novel framework to learn a representation of quantum circuits based on the phase space formulation of many-body systems. Given that $N$-qubit states can be represented as probability distributions on a smooth $2N$-dimensional manifold, we show how unitary gates can be parameterized as normalizing flows between these distributions, opening a new avenue for quantum circuit simulation using generative models. We evaluate our method on three common families of quantum circuits, and show that its accuracy exceeds that of current state-of-the-art neural network methods for large system sizes.
\end{abstract}

\section{Introduction}

The problem of classical simulation of quantum many-body systems is pertinent to both the existence of quantum advantage, and to inference and control of quantum hardware \cite{huang2025vast, nielsen2010quantum}. Naive methods for classical simulation of quantum systems include exact state-vector methods and Feynman-style path integral schemes \cite{penney2017quantum,dawson2004quantum,rudiak607151sum}, which are simple to implement yet have an algorithmic complexity that grows exponentially in the size of the system. Other techniques exploit the structure in certain classes of quantum circuits, such as Clifford-ness \cite{gottesman1998heisenberg} or low tree-rank \cite{markov2008simulating,yoran2008efficientlycontractablequantumcircuits}, which lead to efficient classical algorithms but are not applicable to other circuits. For condensed matter systems in 1D or 2D, tensor network approaches comprise the vast majority of state-of-the-art algorithms \cite{vidal2003efficient,vidal2008class}, under the assumption that the amount of entanglement in the system is bounded by the area of any bipartition. Recently, a new class of algorithms called Pauli propagation \cite{rudolph2025paulipropagationcomputationalframework,rudolph2023classical} has also been shown to approximately simulate quantum circuits in low magic or noisy settings.

Machine-learning techniques have also been applied to the simulation of quantum systems. Neural quantum states (NQS) and its variations \cite{carrasquilla2019reconstructing,carleo2017solving,sharir2020deep,donatella2023dynamics,gutierrez2022real} offer compact representations and were shown to be able to evolve nontrivial many-body states. More recently, generative models have been used to learn representations of quantum states in the probabilistic formulation of quantum dynamics \cite{carrasquilla2021probabilistic,kiktenko2020probability,van_de_Wetering_2018}. However, the accuracy of all of these methods typically degrades with system size and evolution time.

In this paper, we present a novel method to learn the output state of a quantum circuit based on the phase space formalism of many-body qubit systems \cite{heightman2025quantum}. Each intermediate state is represented as a probability distribution, and unitary gates are represented as normalizing flow transformations that map from one distribution to the next. Unlike previous methods, ours is based on a continuous input space rather than a discrete domain \cite{carrasquilla2021probabilistic,carrasquilla2019reconstructing,carleo2017solving,sharir2020deep}, which avoids having to use continuous relaxations or hand-crafted embeddings. 
In addition, our architecture is naturally composable due to the sequential nature of normalizing flows, mirroring the structure of the quantum circuit. This is in contrast with previous methods which retrain a fixed architecture variationally after each gate \cite{donatella2023dynamics,gutierrez2022real,carrasquilla2021probabilistic}. 
We demonstrate our method on three common families of quantum circuits, and show that the classical fidelity of our model exceeds that of two neural network baselines \cite{carleo2017solving,carrasquilla2021probabilistic} as the size and depth of the circuit grows.

\section{Methods}

In this section we describe our framework to learn a representation of the states in a given quantum circuit. We give a brief overview of the phase space of multi-qubit systems before describing how quantum circuits can be modeled using normalizing flows. 

\subsection{Phase space representation}

In the phase-space formalism, $N$-qubit states are represented by probability distributions over the manifold $(S^2)^N$, and unitary maps between states are given by a convolution with an integral kernel (or Feynman-like propagators). Although many different phase space representations exist \cite{heightman2025quantum}, the clear choice here is the $Q$ function due to its positivity, and thus its interpretation as a probability distribution. For an $N$-qubit state $\ket\psi$, the $Q$ function of $\ket\psi$ is given by
\begin{align}
Q_{\ket\psi}(\boldsymbol{\Omega})=\frac{1}{(2\pi)^N}\left|\braket{\boldsymbol{\Omega}}{\psi}\right|^2\label{eq:q_func}
\end{align}
where $\boldsymbol{\Omega}$ is shorthand for $(\theta_1,\varphi_1,\theta_2,\varphi_2,\cdots,\theta_N,\varphi_N)$ specifying a point on $(S^2)^N$, and the spin coherent states $\ket{\boldsymbol{\Omega}}$  are defined as \cite{perelomov1972coherent,radcliffe1971some,zhang1990coherent}
\begin{align}
\ket{\boldsymbol{\Omega}}=\bigotimes_{i=1}^N\ket{\Omega_i}=\bigotimes_{i=1}^N\left(\cos\frac{\theta_i}{2}\ket{0}+e^{i\varphi_i}\sin\frac{\theta_i}{2}\ket{1}\right)
\end{align}
One can show that the $Q$ function is a valid probability distribution over $(S^2)^N$ under the usual spherical measure \cite{heightman2025quantum} . Examples of $Q$ functions for some selected states are shown in App.~\ref{app:q_funcs}, see also \cite{heightman2025quantum}.

Now suppose we have a unitary operator $U$. Since $U$ is linear, there exists some transition kernel $K_U(\boldsymbol{\Omega'},\boldsymbol{\Omega})$ \cite{heightman2025quantum} such that, for any two states $\ket\psi,\ket{\psi'}$ with $\ket{\psi'}=U\ket\psi$, we have
\begin{align}
Q_{\ket{\psi'}}(\boldsymbol{\Omega'})=\int_{(S^2)^N}K_U(\boldsymbol{\Omega'},\boldsymbol{\Omega})\,Q_{\ket{\psi}}(\boldsymbol{\Omega})\,d\boldsymbol{\Omega}\label{eq:q_propagator}
\end{align}
so that unitary maps in the Hilbert space correspond to convolution with the integral kernels $K_U$. Explicitly, given $U$, the kernel $K_U$ has the form
\begin{align}
K_U(\boldsymbol{\Omega'},\boldsymbol{\Omega})&=\frac{1}{(2\pi)^N}\text{Tr}\left[U^\dagger\ketbra{\boldsymbol{\Omega'}}{\boldsymbol{\Omega'}}U \Delta(\boldsymbol{\Omega})\right]&
\Delta(\boldsymbol{\Omega})&=\bigotimes_{i=1}^N\left(3\ketbra{\Omega_i}{\Omega_i}-I\right)
\end{align}
Importantly, if $U$ only acts on a small subset $S$ of qubits, the integral (\ref{eq:q_propagator}) need only be performed over the coordinates $\Omega_i$ with $i\in S$ (since the non-support qubits can be integrated out analytically), so the integral can be computed efficiently in this case. Examples of transition kernels for some common unitaries $U$ are shown in App.~\ref{app:kernel}.

\subsection{Quantum circuits as normalizing flows}

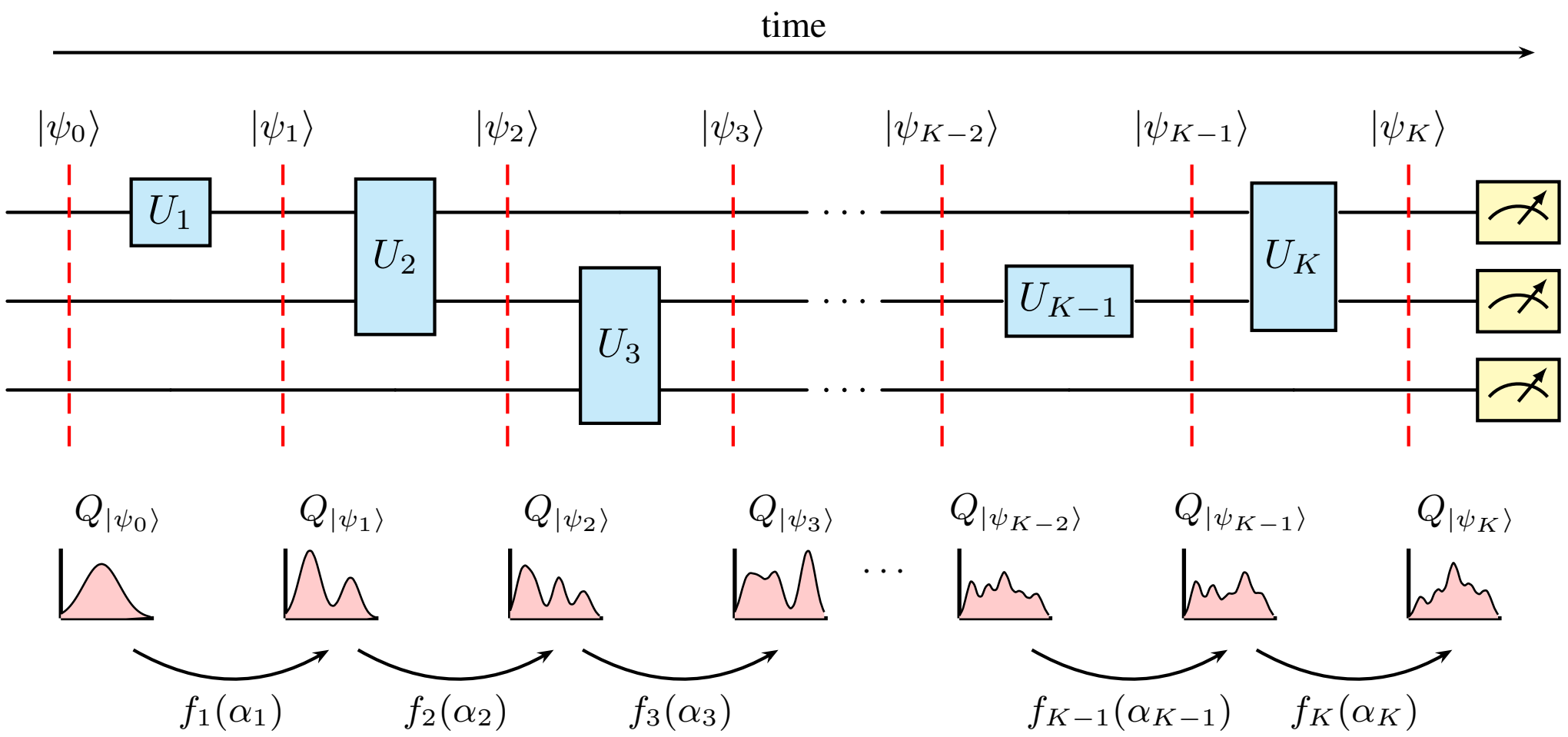
\begin{figure}
\centering
\begin{tikzpicture}
\draw[] (0, 0) node (left) {};
\draw[] (13, 0) node (right) {};
\draw (left) edge[draw,-{Stealth[scale=0.6]},thick] node[above]{\small time} (right);
\draw[] (0, -0.2) node {};  
\end{tikzpicture}
\begin{quantikz}[row sep=5pt, column sep={55pt, between origins}]
\slice{\small $|\psi_0\rangle$} &[-15pt] \gate[style={fill=cyan!20,inner xsep=0pt}]{U_1}\slice{\small $|\psi_1\rangle$} & \gate[2, style={fill=cyan!20,inner xsep=0pt}]{U_2}\slice{\small $|\psi_2\rangle$} & &\,\, \ldots \,\, & & \gate[2, style={fill=cyan!20,inner sep=-1pt}]{U_K}\slice{\small $|\psi_K\rangle$} & \meter[style={fill=yellow!30}]{}\\
&[-15pt]  &  & \gate[2, style={fill=cyan!20,inner xsep=0pt}]{U_3}\slice{\small $|\psi_3\rangle$} & \,\, \ldots \slice{\small $|\psi_{K-2}\rangle$}\,\, & \gate[style={fill=cyan!20,inner xsep=-1pt}]{U_{K-1}}\slice{\small $|\psi_{K-1}\rangle$} & & \meter[style={fill=yellow!30}]{}\\
&[-15pt]  &  & & \,\, \ldots \,\, & & & \meter[style={fill=yellow!30}]{}
\end{quantikz}
\vspace{5pt}
\newline

\begin{tikzpicture}[every circle node/.style={draw, inner sep=-1.2pt, fill=green!20}]
  \draw[] (0.5, 0.9) node {\small $Q_{\ket{\psi_0}}$};
  \draw[line width=1pt] (0, 0) -- (0.8, 0) node {};
  \draw[line width=1pt] (0, 0) -- (0, 0.6) node{};
  \fill [red!20, domain=0:1.1, variable=\x, scale=0.7]
      (0, 0)
      -- plot ({\x}, {exp(-10 * (\x-0.5)^2) / 1.5})
      -- (0.8, 0)
      -- cycle;
  \draw[line width=0.5pt, scale=0.7, domain=0:1.1, smooth, variable=\x] plot ({\x}, {exp(-10 * (\x-0.5)^2) / 1.5});
  \draw[] (0.5, -0.2) node (label0) {};

  \begin{scope}[xshift=55pt]
  \draw[] (0.5, 0.9) node {\small $Q_{\ket{\psi_1}}$};
    \draw[line width=1pt] (0, 0) -- (0.8, 0) node {};
  \draw[line width=1pt] (0, 0) -- (0, 0.6) node{};
  \fill [red!20, domain=0:1.1, variable=\x, scale=0.7]
      (0, 0)
      -- plot ({\x}, {exp(-30 * (\x-0.3)^2) / 1.2 + exp(-40 * (\x-0.8)^2) / 2})
      -- (1, 0)
      -- cycle;
  \draw[line width=0.5pt, scale=0.7, domain=0:1.1, smooth, variable=\x] plot ({\x}, {exp(-30 * (\x-0.3)^2) / 1.2 + exp(-40 * (\x-0.8)^2) / 2});
  \draw[] (0.5, -0.2) node (label1) {};
  \draw (label0) edge[draw,-{Stealth[scale=0.6]},out=330, in=210, looseness=1, thick] node[below]{\small $f_1(\alpha_1)$} (label1);
  \end{scope}

  \begin{scope}[xshift=110pt]
  \draw[] (0.5, 0.9) node {\small $Q_{\ket{\psi_2}}$};
    \draw[line width=1pt] (0, 0) -- (0.8, 0) node {};
  \draw[line width=1pt] (0, 0) -- (0, 0.6) node{};
  \fill [red!20, domain=0:1.1, variable=\x, scale=0.7]
      (0, 0)
      -- plot ({\x}, {exp(-80 * (\x-0.15)^2) / 1.8 + exp(-80 * (\x-0.3)^2) / 2.4 + exp(-80 * (\x-0.6)^2) / 2 + exp(-60 * (\x-0.9)^2) / 3})
      -- (1.1, 0)
      -- cycle;
  \draw[line width=0.5pt, scale=0.7, domain=0:1.1, smooth, variable=\x] plot ({\x}, {exp(-80 * (\x-0.15)^2) / 1.8 + exp(-80 * (\x-0.3)^2) / 2.4 + exp(-80 * (\x-0.6)^2) / 2 + exp(-60 * (\x-0.9)^2) / 3});
  \draw[] (0.5, -0.2) node (label2) {};
  \draw (label1) edge[draw,-{Stealth[scale=0.6]},out=330, in=210, looseness=1, thick]  node[below]{\small $f_2(\alpha_2)$} (label2) ;
  \end{scope}

  \begin{scope}[xshift=165pt]
  \draw[] (0.5, 0.9) node {\small $Q_{\ket{\psi_3}}$};
    \draw[line width=1pt] (0, 0) -- (0.8, 0) node {};
  \draw[line width=1pt] (0, 0) -- (0, 0.6) node{};
  \fill [red!20, domain=0:1.1, variable=\x, scale=0.7]
      (0, 0)
      -- plot ({\x}, {exp(-80 * (\x-0.15)^2) / 2.2 + exp(-80 * (\x-0.3)^2) / 2.5 + exp(-60 * (\x-0.5)^2) / 1.8 + exp(-60 * (\x-0.9)^2) / 1.2})
      -- (1.1, 0)
      -- cycle;
  \draw[line width=0.5pt, scale=0.7, domain=0:1.1, smooth, variable=\x] plot ({\x}, {exp(-80 * (\x-0.15)^2) / 2.2 + exp(-80 * (\x-0.3)^2) / 2.5 + exp(-60 * (\x-0.5)^2) / 1.8 + exp(-60 * (\x-0.9)^2) / 1.2});
  \draw[] (0.5, -0.2) node (label3) {};
  \draw (label2) edge[draw,-{Stealth[scale=0.6]},out=330, in=210, looseness=1, thick]  node[below]{\small $f_3(\alpha_3)$} (label3) ;
  \end{scope}

 \begin{scope}[xshift=188pt]
  \draw[] (0.5, 0.4) node {\small $\cdots$};
  \end{scope}

  \begin{scope}[xshift=220pt]
  \draw[] (0.5, 0.9) node {\small $Q_{\ket{\psi_{K-2}}}$};
    \draw[line width=1pt] (0, 0) -- (0.8, 0) node {};
  \draw[line width=1pt] (0, 0) -- (0, 0.6) node{};
  \fill [red!20, domain=0:1.1, variable=\x, scale=0.7]
      (0, 0)
      -- plot ({\x}, {exp(-120 * (\x-0.15)^2) / 2.2 + exp(-120 * (\x-0.35)^2) / 2.5 + exp(-90 * (\x-0.55)^2) / 1.8 + exp(-90 * (\x-0.75)^2) / 3.2 + exp(-90 * (\x-0.95)^2) / 3.4})
      -- (1.1, 0)
      -- cycle;
  \draw[line width=0.5pt, scale=0.7, domain=0:1.1, smooth, variable=\x] plot ({\x}, {exp(-120 * (\x-0.15)^2) / 2.2 + exp(-120 * (\x-0.35)^2) / 2.5 + exp(-90 * (\x-0.55)^2) / 1.8 + exp(-90 * (\x-0.75)^2) / 3.2 + exp(-90 * (\x-0.95)^2) / 3.4} );
  \draw[] (0.5, -0.2) node (labelk2) {};
  \end{scope}

  \begin{scope}[xshift=275pt]
  \draw[] (0.5, 0.9) node {\small $Q_{\ket{\psi_{K-1}}}$};
    \draw[line width=1pt] (0, 0) -- (0.8, 0) node {};
  \draw[line width=1pt] (0, 0) -- (0, 0.6) node{};
  \fill [red!20, domain=0:1.1, variable=\x, scale=0.7]
      (0, 0)
      -- plot ({\x}, {exp(-120 * (\x-0.15)^2) / 2.2 + exp(-120 * (\x-0.35)^2) / 2.5 + exp(-90 * (\x-0.55)^2) / 3.5 + exp(-90 * (\x-0.75)^2) / 1.8 + exp(-90 * (\x-0.95)^2) / 3.4})
      -- (1.1, 0)
      -- cycle;
  \draw[line width=0.5pt, scale=0.7, domain=0:1.1, smooth, variable=\x] plot ({\x}, {exp(-120 * (\x-0.15)^2) / 2.2 + exp(-120 * (\x-0.35)^2) / 2.5 + exp(-90 * (\x-0.55)^2) / 3.5 + exp(-90 * (\x-0.75)^2) / 1.8 + exp(-90 * (\x-0.95)^2) / 3.4});
  \draw[] (0.5, -0.2) node (labelk1) {};
  \draw (labelk2) edge[draw,-{Stealth[scale=0.6]},out=330, in=210, looseness=1, thick]  node[below]{\small $f_{K-1}(\alpha_{K-1})$} (labelk1) ;
  \end{scope}

  \begin{scope}[xshift=330pt]
  \draw[] (0.5, 0.9) node {\small $Q_{\ket{\psi_K}}$};
    \draw[line width=1pt] (0, 0) -- (0.8, 0) node {};
  \draw[line width=1pt] (0, 0) -- (0, 0.6) node{};
  \fill [red!20, domain=0:1.1, variable=\x, scale=0.7]
      (0, 0)
      -- plot ({\x}, {exp(-120 * (\x-0.15)^2) / 3.8 + exp(-120 * (\x-0.35)^2) / 3 + exp(-90 * (\x-0.55)^2) / 1.5 + exp(-90 * (\x-0.75)^2) / 2.5 + exp(-90 * (\x-0.95)^2) / 3})
      -- (1.1, 0)
      -- cycle;
  \draw[line width=0.5pt, scale=0.7, domain=0:1.1, smooth, variable=\x] plot ({\x}, {exp(-120 * (\x-0.15)^2) / 3.8 + exp(-120 * (\x-0.35)^2) / 3 + exp(-90 * (\x-0.55)^2) / 1.5 + exp(-90 * (\x-0.75)^2) / 2.5 + exp(-90 * (\x-0.95)^2) / 3});
  \draw[] (0.5, -0.2) node (labelk) {};
  \draw (labelk1) edge[draw,-{Stealth[scale=0.6]},out=330, in=210, looseness=1, thick]  node[below]{\small $f_{K}(\alpha_K)$} (labelk) ;
  \end{scope}

\end{tikzpicture}
\caption{Our proposed framework for learning the output state of a quantum circuit. In the unitary picture (top half), the state of a quantum circuit is represented by the state vectors $\ket{\psi_k}$, and maps from one state to the next are given by unitary operators $U_k$. In the phase-space picture (bottom half), the state of a quantum circuit is represented by its $Q$ function, and maps from one state to the next are induced by flow transformations $f_k$ of phase space, which we parameterize by $\alpha_k$. Here, the distributions are shown over $\mathbb{R}$ for illustrative purposes, but in reality are over $(S^2)^N$.}
\label{fig:framework}
\vspace{-5pt}
\end{figure}


A quantum circuit is specified by an initial state $\ket{\psi_0}$ along with an ordered collection of gates $U_1,U_2,\cdots,U_K$. The state after the $k$-th gate is given by 
\begin{align}
\ket{\psi_k}=U_kU_{k-1}\cdots U_2U_1\ket{\psi_0}
\end{align}

From the previous section, each intermediate state $\ket{\psi_k}$ can be represented as a probability distribution over $(S^2)^N$ via Eq.~(\ref{eq:q_func}), and mapping between states $\ket{\psi_{k-1}}$ and $\ket{\psi_{k}}$ is accomplished by substituting the corresponding gate $U_k$ into Eq.~(\ref{eq:q_propagator}).  However, $K_{U_k}$ may in general contain negative values, which renders the naive efficient sampling scheme from the distribution $Q_{\ket{\psi_{k}}}(\boldsymbol{\Omega'})$ given $Q_{\ket{\psi_{k-1}}}(\boldsymbol{\Omega})$ inapplicable \cite{van_de_Wetering_2018,chruscinski2015pseudo}. Our solution is therefore to represent the $Q$ function of $\ket{\psi_{k}}$ as a normalizing flow using the $Q$ function of $\ket{\psi_{k-1}}$ as the initial distribution. More specifically, suppose that $f_k: (S^2)^N\to(S^2)^N$ is a flow transformation corresponding to $U_k$. Then our goal is to find $f_k$ such that the $Q$ function of $\ket{\psi_k}$ is well-approximated:
\begin{align}
Q_{\ket{\psi_k}}(\boldsymbol{\Omega})\approx \tilde{Q}_{\ket{\psi_k}}(\boldsymbol{\Omega})= Q_{\ket{\psi_{k-1}}}\left(f_k^{-1}(\boldsymbol{\Omega})\right)\left|\det\frac{\partial f_k^{-1}}{\partial\boldsymbol{\Omega}}\right| 
\end{align}
with the right-hand-side being the usual change-of-variables formula for the probability density \cite{rezende2015variational}. This process is composable: to approximate the $Q$ function for a state $\ket{\psi_k}$, we can pushforward the $Q$ function for the initial state $\ket{\psi_0}$ through each of the flows $f_1,f_2,\cdots,f_k$ in order.

We emphasize that the $Q$ function is a distribution over an $2N$-dimensional manifold, and thus the normalizing flows can be implemented efficiently. The problem of the exponential growth of the Hilbert space is now a problem of model expressivity \cite{carleo2017solving,carrasquilla2019reconstructing}.

The initial state $\ket{\psi_0}$ is usually some fiducial state, which can be taken to be $\ket{00\cdots 0}$. This has the simple product $Q$ function given by (see also App.~\ref{app:q_funcs}):
\begin{align}
Q_{\ket{\psi_0}}(\boldsymbol{\Omega})=\prod_{i=1}^N Q_{\ket{0}}(\Omega_i)=\frac{1}{(4\pi)^N}\prod_{i=1}^N(1+\cos\theta_i)
\end{align}
where, importantly, sampling and computing likelihoods are both efficient and exact. We parameterize each flow $f_k$ with trainable parameters $\alpha_k$, and train to minimize the KL divergence between the $Q$ functions obtained via the transition kernel (\ref{eq:q_propagator}) and the ones given by the trainable flows:
\begin{align}
\mathcal{L}(\boldsymbol{\alpha})=\sum_{k=1}^K \mathcal{D}\left(\int K_{U_k}(\boldsymbol{\Omega'},\boldsymbol{\Omega})\,\tilde{Q}_{\ket{\psi_{k-1}}}(\boldsymbol{\Omega})\,d\boldsymbol{\Omega}\,\,\bigg\|\,\,\tilde{Q}_{\ket{\psi_{k}}}\left(\boldsymbol{\Omega};\alpha_k\right)\right)\label{eq:loss_fn}
\end{align}
where $\tilde{Q}_{\ket{\psi_{k}}}$ is the model's approximation of $Q_{|\psi_k\rangle}$, and $\tilde Q_{\ket{\psi_0}}=Q_{|\psi_0\rangle}$. Assuming each gate $U_k$ acts on a small number of qubits, the KL divergences can be estimated efficiently using Monte Carlo sampling; the $k$-th term requires samples from the $k-1$-th $Q$ function, which can be obtained efficiently due to the composability of normalizing flows. Thus training uses only the learned $Q$ functions, and the exact $Q$ functions are used only to evaluate the trained model. 

In practice, we pre-train each of the flows sequentially from $k=1$ to $k=K$ for better stability in the early stages of training, and then fine-tune all flows jointly on Eq.~\ref{eq:loss_fn}. Our framework is illustrated in Fig.~\ref{fig:framework}.

\subsection{Related work}

Normalizing flows have also been applied to quantum systems directly. Closest to our work is Q-Flow \cite{dugan2023qflow}, which likewise represents the Husimi $Q$ function with a normalizing flow, but for open continuous-variable systems: the phase space is $\mathbb{R}^{2N}$, and the flow is evolved in continuous time under the PDE that the Lindblad dynamics induces for $Q$. Flows have also been used as neural quantum states for continuous-variable many-body systems~\cite{lawrence2024}, and flow matching on spin Wigner functions has been used to learn distributions over multi-qubit states~\cite{hahm2024}. Our setting differs in three ways. First, we treat qubits, whose phase space is the compact manifold $(S^2)^N$. Second, the dynamics is a sequence of discrete gates, whose phase-space action is a convolution rather than a differential operator. Third, instead of re-fitting or evolving a single flow, we assign one flow per gate and compose them.

\section{Experiments}

\begin{figure}
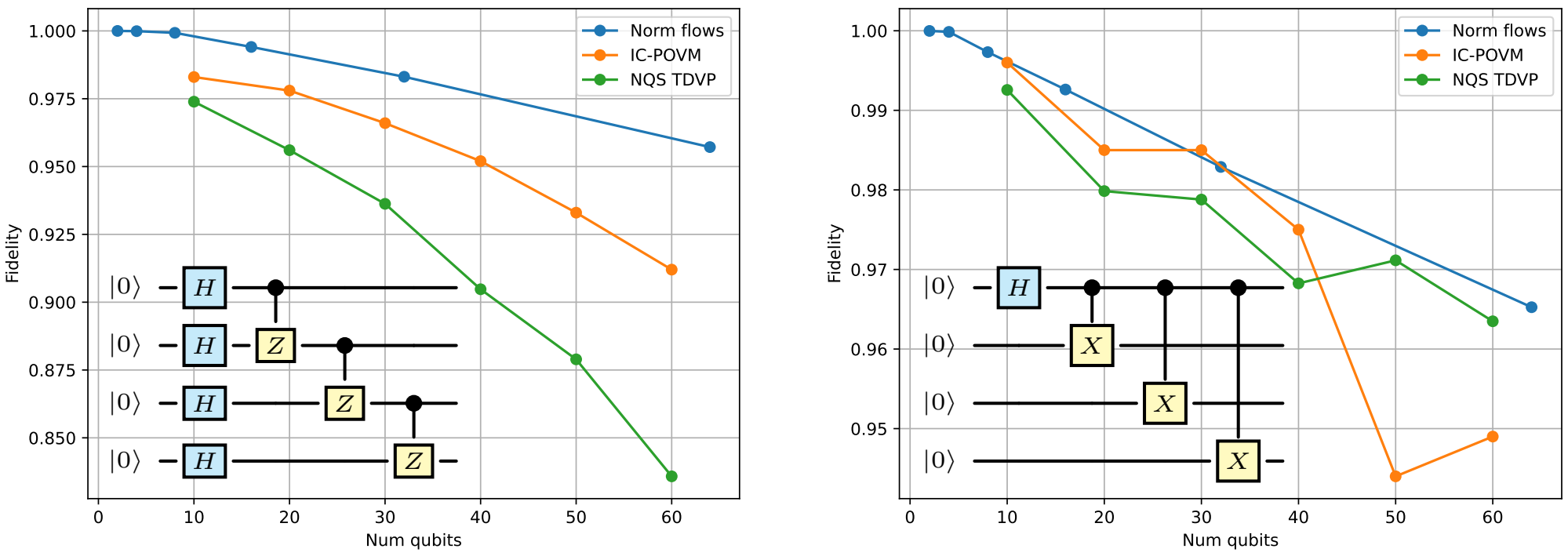

    \captionsetup[subfigure]{aboveskip=0pt,belowskip=0pt}
    \centering
    \begin{subfigure}{0.49\textwidth}
        \centering
        \input{Figures/cluster.tikz}
        \caption{1D cluster state preparation}
    \end{subfigure}
    \begin{subfigure}{0.49\textwidth}
        \centering
        \input{Figures/ghz.tikz}
        \caption{GHZ state preparation}
    \end{subfigure}
    \caption{Classical fidelity between the exact $Q$ representation of the final state and the output distribution of the normalizing flow model, as a function of the total number of qubits. For comparison, we show the fidelities obtained in \cite{carleo2017solving} and \cite{carrasquilla2021probabilistic}.}
    \label{fig:fidelity_plots}
    \vspace{-5pt}
\end{figure}

\begin{figure}
    \centering
    \includegraphics[width=0.80\textwidth]{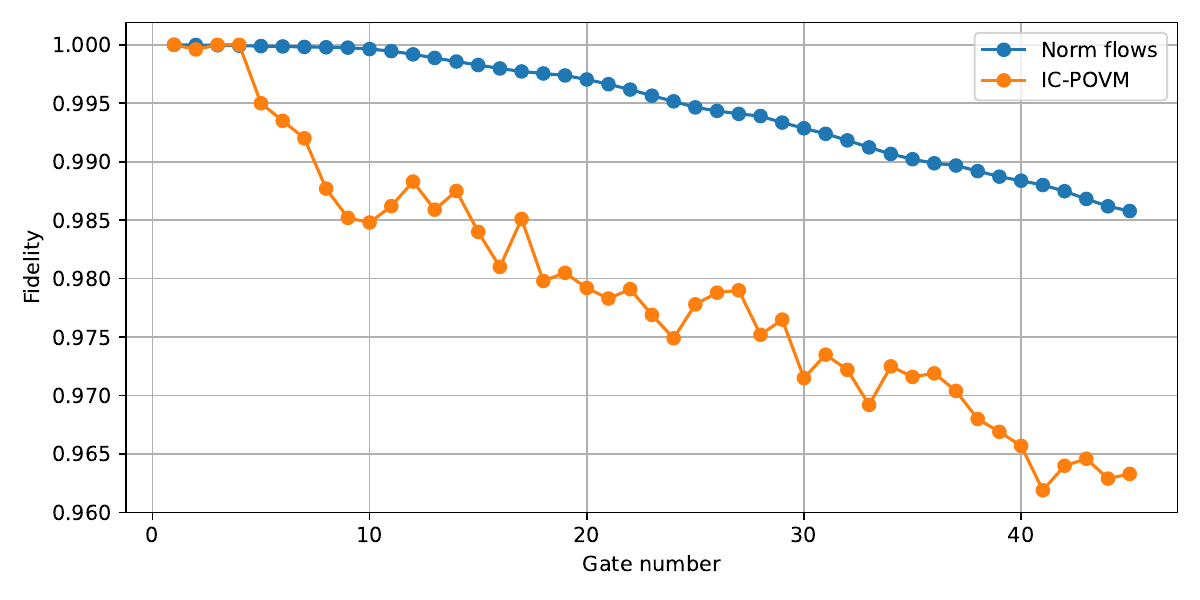}
        \caption{Classical fidelity between the exact $Q$ representation of each intermediate state and the learned distribution of the normalizing flow model, as a function of the gate number for a 10-qubit VQE circuit for TFIM ground state preparation. For comparison, we show the fidelities obtained in \cite{carrasquilla2021probabilistic}.}
    \label{fig:vqe_fid}
    \vspace{-5pt}
\end{figure}

To evaluate our framework, we train normalizing flow models for three families of quantum circuits: 1D cluster state preparation, GHZ state preparation, and ground state preparation of a 1D transverse-field Ising model (TFIM) at the critical point. These are the same families of circuits that were used for evaluation in \cite{carrasquilla2021probabilistic}. 

For our normalizing flow architecture, we use a continuous normalizing flow \cite{grathwohl2018ffjord,falorsi2021continuous} over the manifold $(S^2)^N$ for each unitary operator, using a simple multi-layer perceptron (MLP) to parameterize the vector field.  We train on the loss function in equation (\ref{eq:loss_fn}) with the Adam optimizer \cite{kingma2017adammethodstochasticoptimization} with learning rate 0.001 until convergence.

Nothing in our method uses the Clifford structure of the GHZ and cluster circuits. They were chosen because exact $Q$ functions are available at large $N$, and to allow direct comparison with \cite{carrasquilla2021probabilistic}. In contrast, the VQE circuit contains arbitrary-angle rotations and is not Clifford.

We evaluate the trained model by computing the classical fidelity between the $Q$ function of the true state $Q_\ket{\psi_K}$ and the final predicted distribution $\tilde{Q}$:
\begin{align}
F\left(Q_\ket{\psi_K},\tilde{Q}\right)= \int_{(S^2)^N}\sqrt{Q_\ket{\psi_K}(\boldsymbol{\Omega})\tilde{Q}(\boldsymbol{\Omega})}\,d\boldsymbol{\Omega}\label{eq:clas_fid}
\end{align}
since it is efficient to estimate for large system sizes (whereas computing the quantum fidelity is intractable).
We compare against the results in \cite{carrasquilla2021probabilistic}, which uses a transformer architecture to learn the IC-POVM probability vector of a state instead of the $Q$ function, and evaluates it using the discrete version of the classical fidelity (\ref{eq:clas_fid}). We pull the metrics directly from \cite{carrasquilla2021probabilistic}.
For the 1D cluster state and GHZ state, we also benchmark against neural quantum states (NQS) \cite{carleo2017solving}, which uses the TDVP algorithm to find optimal parameters that solve the dynamics. In this case, we run the training routine from scratch, using an MLP to parameterize the real and imaginary parts of the wavefunction and the stochastic reconfiguration algorithm for TDVP time-stepping.

The results are shown in Figures \ref{fig:fidelity_plots} and \ref{fig:vqe_fid}. At the largest sizes, our model reaches $F\approx0.96$ at $N=64$ for the cluster state, compared with $0.91$ (IC-POVM) and $0.84$ (NQS) at $N=60$, and $F\approx0.965$ for the GHZ state, compared with $0.95$ (IC-POVM) and $0.96$ (NQS). For the VQE circuit, we see that the fidelity decreases roughly linearly by $\sim 3\times10^{-4}$ per gate, reaching $0.985$ compared with $0.965$ for the transformer.

\section{Discussion and limitations}\label{sec:discussion}

\paragraph{Computational cost.} Each gate has its own flow, so the number of parameters grows linearly with the number of gates $K$.  Training gate $k$ requires evaluating $\tilde Q_{k-1}$ at new points, which means inverting the $k-1$ preceding flows, so the total training cost grows as $\mathcal{O}(K^2)$. In contrast, the baseline neural methods retrain a fixed-size model after each gate, so their total training cost grows as $\mathcal{O}(K)$. As a result, Figures \ref{fig:fidelity_plots} and \ref{fig:vqe_fid} compare accuracy at a given system size and are not compute-matched.

\paragraph{Negativity.} The model is necessary in the first place because the transition kernels $K_{U_k}$ generally contain negative values. Despite this, convolution with the kernel can be computed efficiently assuming that $U_k$ only has support on a small number of qubits (say 2), since the unsupported coordinates can be integrated out analytically. On the other hand, simulating the whole circuit as a single unitary gate would require integrating over all $2N$ coordinates. Thus the sign problem manifests as approximation error of the learned model, which compounds exponentially with the circuit depth (see Figure \ref{fig:vqe_fid}).

\paragraph{State dependence.} For a single-qubit gate $U$ acting on qubit $i$, the $Q$ function transforms exactly via the rotation $R_U$ that $U$ induces on the $i$-th Bloch sphere. Hence $f_k=R_{U_k}$ is an exact flow that is independent of the input, including for non-Clifford gates such as $T$. For entangling gates however, no such flow exists in general. Each learned $f_k$ is therefore an approximate transport map from $\tilde Q_{k-1}$ to its convolution with $K_{U_k}$. For a different input state $\hat Q_{k-1}$, the transport map does not carry over and would need to be retrained.

\section{Conclusion and outlook}

In this contribution, we introduced a novel framework to learn the output state of a quantum circuit based on normalizing flows in the phase space representation, and showed numerically that the method is able to scale well to large system sizes. 

Natural next benchmarks are Trotterized dynamics $e^{-iHt}$ of local Hamiltonians, in regimes that challenge tensor networks and Pauli propagation, as well as noisy and open-system evolution \cite{lindblad1976generators}. Since local observables can be estimated from flow samples, the composed flow could also serve as a variational ansatz: one would optimize the gate parameters of a circuit using energies estimated through the flows. 

\section{Acknowledgements}

The authors acknowledge support from Quside Technologies SL for providing GPU resources used in the experiments.

EJ acknowledges support from the predoctoral program AGAUR-FI ajuts (2025
FI-1 00298) Joan Oró, which is backed by the Secretariat of Universities and
Research of the Department of Research and Universities of the Generalitat
of Catalonia, as well as the European Social Plus Fund.

TH acknowledges support from the Government of Spain (Severo Ochoa CEX2019-000910-S, Quantum in Spain, FUNQIP and European Union NextGenerationEU PRTR-C17.I1), the European Union (PASQuanS2.1, 101113690 and Quantera Veriqtas), Fundació Cellex, Fundació Mir-Puig, Generalitat de Catalunya (CERCA program), the ERC AdG CERQUTE and the AXA Chair in Quantum Information Science.

{
\small
\bibliographystyle{unsrt}
\bibliography{references2026}
}

\appendix

\newpage

\section{$Q$ Functions of Selected States}\label{app:q_funcs}

Recall that the $Q$ function is defined as
\begin{align}
Q_{\ket\psi}(\boldsymbol{\Omega})=\frac{1}{(2\pi)^N}\left|\braket{\boldsymbol{\Omega}}{\psi}\right|^2
\end{align}
where $\boldsymbol{\Omega}=(\theta_1,\varphi_1,\theta_2,\varphi_2,\cdots,\theta_N,\varphi_N)$, and the coherent state $\ket{\boldsymbol{\Omega}}$ is given by
\begin{align}
\ket{\boldsymbol{\Omega}}=\left(\cos\frac{\theta_1}{2}\ket{0}+e^{i\varphi_1}\sin\frac{\theta_1}{2}\ket{1}\right)\otimes\cdots\otimes \left(\cos\frac{\theta_N}{2}\ket{0}+e^{i\varphi_N}\sin\frac{\theta_N}{2}\ket{1}\right)
\end{align}

In Table \ref{tab:q_funcs} we show the $Q$ functions for the eigenstates of the single-qubit Pauli operators, the four 2-qubit Bell states, and the 3-qubit GHZ and W states.

\begingroup
\renewcommand{\arraystretch}{2.4} 
\begin{table}[h]
    \small
  \caption{$Q$ functions for selected states}
  \label{tab:q_funcs}
  \centering
  \begin{tabular}{ccc}
    \toprule
     \# qubits & State  & $Q$ function   \\
    \midrule
    1 & $\ket{0}$ & $\frac{1}{4\pi}\left(1+\cos\theta\right)$ \\ 
    1 & $\ket{1}$ & $\frac{1}{4\pi}\left(1-\cos\theta\right)$ \\ 
    1 & $\ket{+}$ & $\frac{1}{4\pi}\left(1+\sin\theta\cos\varphi\right)$ \\ 
    1 & $\ket{-}$ & $\frac{1}{4\pi}\left(1-\sin\theta\cos\varphi\right)$ \\ 
    1 & $\ket{{+}i}$ & $\frac{1}{4\pi}\left(1+\sin\theta\sin\varphi\right)$ \\ 
    1 & $\ket{{-}i}$ & $\frac{1}{4\pi}\left(1-\sin\theta\sin\varphi\right)$ \\ 
    2 & $\ket{\Phi^+}$ & $\frac{1}{(4\pi)^2}\left(1+\cos\theta_1\cos\theta_2+\sin\theta_1\sin\theta_2\cos(\varphi_1+\varphi_2)\right)$ \\
    2 & $\ket{\Phi^-}$ & $\frac{1}{(4\pi)^2}\left(1+\cos\theta_1\cos\theta_2-\sin\theta_1\sin\theta_2\cos(\varphi_1+\varphi_2)\right)$ \\
    2 & $\ket{\Psi^+}$ & $\frac{1}{(4\pi)^2}\left(1-\cos\theta_1\cos\theta_2+\sin\theta_1\sin\theta_2\cos(\varphi_1+\varphi_2)\right)$ \\
    2 & $\ket{\Psi^-}$ & $\frac{1}{(4\pi)^2}\left(1-\cos\theta_1\cos\theta_2-\sin\theta_1\sin\theta_2\cos(\varphi_1+\varphi_2)\right)$ \\
    \multirow{2}{*}{3} & \multirow{2}{*}{$\ket{\text{GHZ}}$}  & \small $\begin{aligned}[t]&\frac{1}{(4\pi)^3}(1+\cos\theta_1\cos\theta_2+\cos\theta_1\cos\theta_3+\cos\theta_2\cos\theta_3\\&\qquad\qquad+\sin\theta_1\sin\theta_2\sin\theta_3\cos(\varphi_1+\varphi_2+\varphi_3))\end{aligned}$\\
    \multirow{3}{*}{3} & \multirow{3}{*}{$\ket{\text{W}}$}  & \small $\begin{aligned}[t]&\frac{1}{3(4\pi)^3}\Big((1+\cos\theta_1)(1-\cos\theta_2\cos\theta_3+2\sin\theta_2\sin\theta_3\cos(\varphi_2-\varphi_3))\nonumber\\
&\qquad\qquad\qquad+(1+\cos\theta_2)(1-\cos\theta_1\cos\theta_3+2\sin\theta_1\sin\theta_3\cos(\varphi_1-\varphi_3))\nonumber\\
&\qquad\qquad\qquad+(1+\cos\theta_3)(1-\cos\theta_1\cos\theta_2+2\sin\theta_1\sin\theta_2\cos(\varphi_1-\varphi_2))\Big)\end{aligned}$\\
    \bottomrule
  \end{tabular}
\end{table}

\endgroup

\section{Transition Kernels of Selected Unitaries}\label{app:kernel}

Recall that the transition kernel $K_U(\boldsymbol{\Omega'},\boldsymbol{\Omega})$ of a unitary $U$ is defined as
\begin{align}
Q_{\ket{\psi'}}(\boldsymbol{\Omega'})=\int_{(S^2)^N}K_U(\boldsymbol{\Omega'},\boldsymbol{\Omega})\,Q_{\ket{\psi}}(\boldsymbol{\Omega})\,d\boldsymbol{\Omega}
\end{align}
for all states $\ket{\psi},\ket{\psi'}$ such that $\ket{\psi'}=U\ket\psi$. In Table \ref{tab:kernels} we show the kernels for the Pauli $X,Y,Z$ and $H$ gates, as well as an arbitrary $Z$ rotation and the two-qubit controlled $Z$ gate. Altogether these operators form a universal set of gates.

For brevity, we define
\begin{align}
X(\Omega)&=X(\theta,\varphi)=\sin\theta\cos\varphi \\
Y(\Omega)&=Y(\theta,\varphi)=\sin\theta\sin\varphi \\
Z(\Omega)&=Z(\theta,\varphi)=\cos\theta
\end{align}

\begingroup
\renewcommand{\arraystretch}{2.4} 
\begin{table}[h!]
    \small
  \caption{Transition kernels for common unitary operations}
  \label{tab:kernels}
  \centering
  \begin{tabular}{cc}
    \toprule
    Unitary $U$   & Kernel $K_U(\boldsymbol{\Omega'},\boldsymbol{\Omega})$   \\
    \midrule
    $X$ & $\frac{1}{4\pi}\left(1+3X(\Omega')X(\Omega)-3Y(\Omega')Y(\Omega)-3Z(\Omega')Z(\Omega)\right)$ \\
    $Y$ & $\frac{1}{4\pi}\left(1-3X(\Omega')X(\Omega)+3Y(\Omega')Y(\Omega)-3Z(\Omega')Z(\Omega)\right)$ \\
    $Z$ & $\frac{1}{4\pi}\left(1-3X(\Omega')X(\Omega)-3Y(\Omega')Y(\Omega)+3Z(\Omega')Z(\Omega)\right)$ \\
    $H$ & $\frac{1}{4\pi}\big(1+3X(\Omega')Z(\Omega)-3Y(\Omega')Y(\Omega)+3Z(\Omega')X(\Omega)\big)$ \\
    $R_z(\alpha)$ & $\frac{1}{4\pi}\left(1+3X(\theta',\varphi'-\alpha)X(\Omega)+3Y(\theta',\varphi'-\alpha)Y(\Omega)+3Z(\Omega')Z(\Omega)\right)$ \\
    $CZ$ & $\begin{aligned}[t]\tfrac{1}{(4\pi)^2}\Big[&\big(1+3Z(\Omega'_1)Z(\Omega_1)\big)\big(1+3Z(\Omega'_2)Z(\Omega_2)\big)\\
&+3\big(3Z(\Omega_1)+Z(\Omega'_1)\big)\big(X(\Omega'_2)X(\Omega_2)+Y(\Omega'_2)Y(\Omega_2)\big)\\
&+3\big(X(\Omega'_1)X(\Omega_1)+Y(\Omega'_1)Y(\Omega_1)\big)\big(3Z(\Omega_2)+Z(\Omega'_2)\big)\\
&+9\big(Y(\Omega'_1)X(\Omega_1)-X(\Omega'_1)Y(\Omega_1)\big)\big(Y(\Omega'_2)X(\Omega_2)-X(\Omega'_2)Y(\Omega_2)\big)\Big]\end{aligned}$ \\
    \bottomrule
  \end{tabular}
\end{table}

\endgroup

\end{document}